\documentclass[english,review]{elsarticle}
\usepackage[OT1]{fontenc}
\usepackage[latin9]{inputenc}
\usepackage{geometry}
\usepackage{xcolor}
\usepackage{pdfcolmk}
\usepackage{float}
\usepackage{amsmath}
\usepackage{amssymb}
\usepackage{graphicx}
\usepackage{setspace}
\PassOptionsToPackage{normalem}{ulem}
\usepackage{ulem}
\makeatletter

\providecommand{\tabularnewline}{\\}
\providecolor{lyxadded}{rgb}{0,0,1}
\providecolor{lyxdeleted}{rgb}{1,0,0}

\usepackage{nicefrac}
  \usepackage{hyperref}
\usepackage{url}
  \hypersetup{colorlinks=true,citecolor=blue}

\makeatother

\usepackage{babel}
\begin{document}

\title{Numerical Stability of Explicit Off-lattice Boltzmann Schemes: A
comparative study }

\author{Parthib R. Rao\fnref{fn1}}

\ead{prr28@pitt.edu}

\author{Laura A. Schaefer}

\ead{las149@pitt.edu}

\address{Department of Mechanical Engineering and Material Science, University
of Pittsburgh,}

\address{636 Benedum Hall, 3700 O\textquoteright{}Hara St.,}

\address{Pittsburgh PA, 15261, USA.}
\begin{abstract}
The off-lattice Boltzmann (OLB) method consists of numerical schemes
which are used to solve the discrete Boltzmann equation. Unlike the
commonly used lattice Boltzmann method, the spatial and time steps
are uncoupled in the OLB method. In the currently proposed schemes,
which can be broadly classified into Runge-Kutta-based and characteristics-based,
the size of the time-step is limited due to numerical stability constraints.
In this work, we systematically compare the numerical stability of
the proposed schemes in terms of the maximum stable time-step. In
line with the overall LB method, we investigate the available schemes
where the advection approximation is explicit, and the collision approximation
is either explicit or implicit. The comparison is done by implementing
these schemes on benchmark incompressible flow problems such as Taylor
vortex flow, Poiseuille flow and, lid-driven cavity flow. It is found
that the characteristics-based OLB schemes are numerically more stable
than the Runge-Kutta-based schemes. Additionally, we have observed
that, with respect to time-step size, the scheme proposed by Bardow
\emph{et al. \citep{Bardow2006General}} is the most numerically stable
and computationally efficient scheme compared to similar schemes,
for the flow problems tested here. \end{abstract}
\begin{keyword}
Off-lattice Boltzmann method, finite-difference, numerical stability

\fntext[fn1]{Corresponding author, Tel.no. 412-624-9720}
\end{keyword}
\maketitle

\section{Introduction}

The lattice Boltzmann (LB) method is an alternative and powerful numerical
technique used for modeling a variety of complex hydrodynamic flows
\citep{Aidun2010,Succi2001}. Unlike conventional numerical methods
which discretize the macroscale governing equations directly, the
LB method solves a fully-discrete kinetic equation for distribution
functions (DFs) $f_{i}(\boldsymbol{x},t)$, designed to reproduce
the Navier-Stokes equation in the hydrodynamic limit. The LB method
has advantages such as ease of parallelization, simplicity of programming,
and a capability for incorporating model interactions for simulating
complex flows.

A defining feature of the LB method is the coupling between the velocity
and space-time discretizations. That is, for a particular discrete-velocity
set, $\boldsymbol{\xi}_{i}$, the coupling automatically fixes the
temporal and spatial steps through the relation $\Delta\boldsymbol{x}=\boldsymbol{\xi}_{i}\Delta t$.
This procedure has some advantages such as numerical-diffusion free
(exact) advection and computational efficiency (copy-operation). The
coupling is, in fact, a carryover from the earliest LB models, which
were based on Lattice Gas Automata (LGA). However, the LGA link was
broken when it was shown more than a decade ago that the LB method
can be derived directly from the discrete Boltzmann equation as a
special finite-difference scheme \citep{Abe1997Derivation,He1997,Shan2006Kinetic}.
Consequently, the velocity-space can be discretized according to the
flow-physics to be modeled. The discretization of space and time is
a numerical requirement and, importantly, is not tied to the discretization
of the velocity-space.

As a consequence, a subset of the LB method, called the off-lattice
Boltzmann (OLB) method, was developed where space and time are \emph{independently}
discretized, i.e. $\Delta\boldsymbol{x}\neq\boldsymbol{\xi}_{i}\Delta t$.
In the OLB method, we do not have the simplicity of a Lagrangian-type
of evolution (streaming), rather the evolution of $f_{i}$ takes place
in an Eulerian sense. The earliest OLB schemes were geared mainly
towards extending the geometric flexibility of the LB method, which
was previously limited, due to the requirement of a uniform Cartesian
mesh. Several OLB schemes with different spatial discretization methods
such as finite-volume (FV), finite-element (FE), and finite-difference
(FD), along with their variants, have been developed. For example,
OLB schemes were used for non-uniform mesh \citep{Cao1997Physical},
curvilinear co-ordinates \citep{Mei1998OnFD,Guo2003Explicit}, unstructured
mesh \citep{Nannelli1992,Patil2009,Ubertini2003}, finite element
mesh \citep{Lee2001Characteristic,Bardow2006General} among others.
These advancements have made the LB method feasible for many practical
engineering problems. 

In addition to improving the geometric flexibility of the LB method,
OLB schemes can also be used to solve the discrete Boltzmann equation
(DBE) with higher-order lattices. Higher-order lattices are sets of
discrete velocities, which are more suited to model more complex flows
such as thermal flows, micro-scale (high Knudsen number) flows, etc.
In many of these velocity sets (also termed as \emph{non-space-filling
or off-lattice}), the discrete velocities cannot be expressed as an
integer multiple of the smallest non-trivial speed. The D2Q16 velocity-set
listed in \citep{Shan2006Kinetic} and \citep{Chikatamarla2006} and
the D2Q17 velocity-set in \citep{Surmas2009} are typical examples.
Since the regular stream-collide type of evolution scheme cannot be
employed with these lattices, OLB schemes provide a viable evolution
scheme for the DBE. 

While several sophisticated spatial-discretization methods have been
developed, many of the studies use time-marching schemes such as explicit
Euler or Runge-Kutta (RK) for temporal discretization. Typically,
these schemes require very small values of $\Delta t$ relative to
the relaxation parameter $\tau$, to maintain numerical stability
\citep{Ubertini2008Generalised,Xu2003lattice}. This is in contrast
to the LB method, which offers unconditional stability. Small $\Delta t$
requirement is particularly restrictive in the case of flows with
high Reynolds number $(Re)$ flows where $\tau$ is very small. Moreover,
in the LB method, the Mach number \emph{Ma} in the simulations has
to be kept small (generally less than $0.1)$ to limit the compressibility
errors. Small values of \emph{Ma} lead to a slower convergence rate,
especially for steady-state flow problems \citep{Guo2004Preconditioned,Turkel1987}.
Thus, the combined effects of small \emph{Ma} and $\Delta t$ increase
the overall computational cost of the RK-based OLB schemes. 

Many alternative time-marching schemes have been proposed that maintain
the numerical stability of the OLB method at higher values of $\Delta t$,
relative to the relaxation parameter $\tau$, i.e. at higher $\Delta t/\tau$
values. These schemes vary greatly in their numerical stability due
to the different approximations of the collision and advection part
of the DBE. Hence, there is a need to systematically compare their
\emph{relative} performance in terms of the numerical stability of
these schemes. This work addresses this need. 

More specifically, we assess the stability of various OLB schemes,
as quantified in terms of their maximum allowable $\Delta t/\tau$
ratio. This is done via benchmark testing on incompressible flow problems
such as Taylor-vortex flow, Poiseuille flow and lid-driven cavity
flow. \textcolor{blue}{The on-lattice D2Q9 velocity set, which is
used here for evaluation purposes, is described in Section 2.1. The
various time-marching (OLB) schemes used in the comparative analysis
are described in brief in Section 2.2. }

\section{Numerical Formulation}

\subsection{Discrete Boltzmann Equation}

\begin{figure}[H]
\begin{centering}
\includegraphics[scale=0.4]{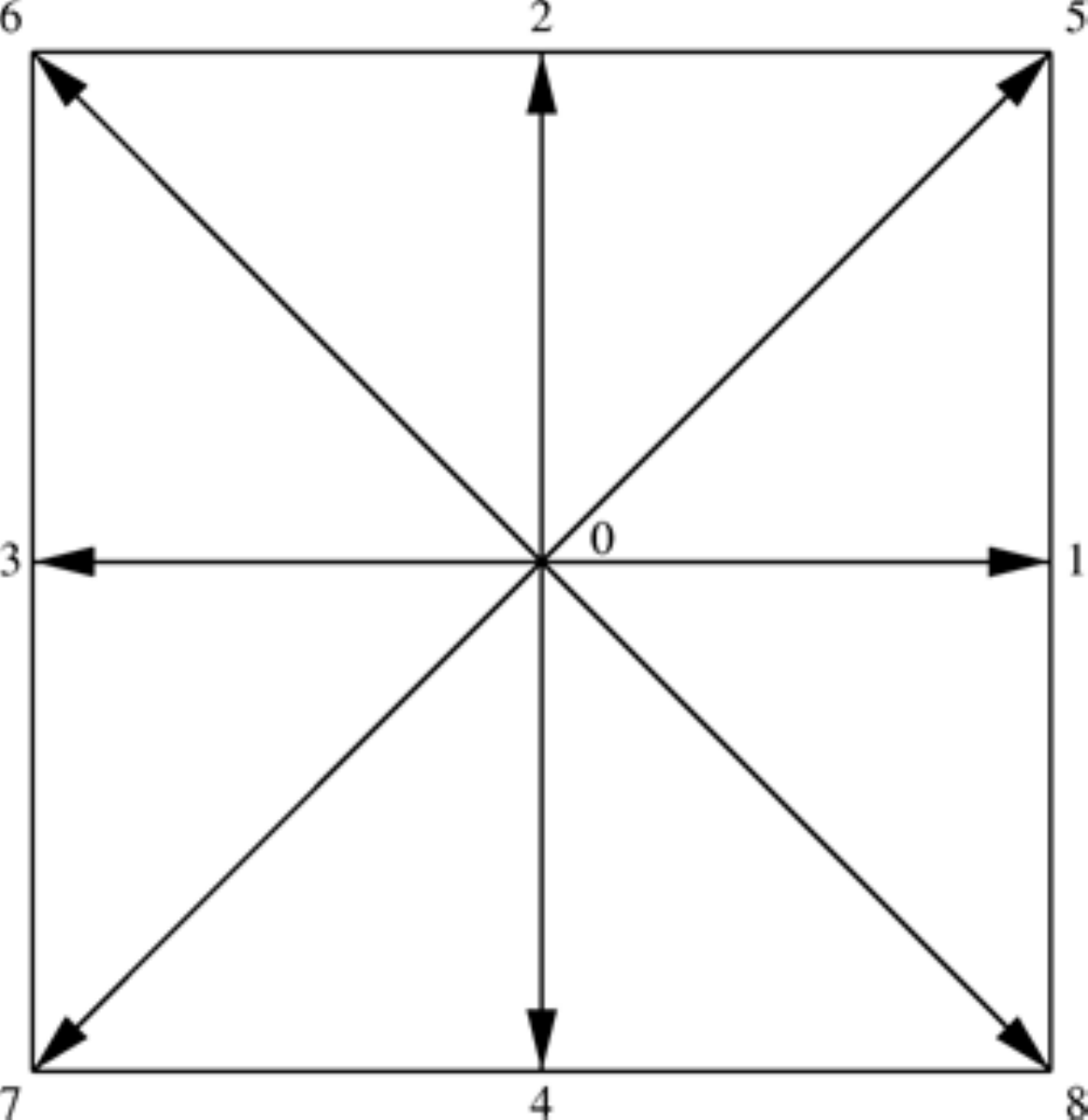}
\par\end{centering}

\caption{D2Q9 velocity lattice.\label{fig:D2Q9}}
\end{figure}

The basis for all OLB schemes is the Boltzmann equation with the Bhatnagar-Gross-Krook
collision approximation \citep{Bhatnagar1954}, which is given as:

\begin{equation}
\frac{\partial f}{\partial t}+\boldsymbol{\xi}\cdot\nabla f=-\frac{1}{\tau}(f-f^{eq}),\label{eq:BBGK}
\end{equation}
where $f\equiv f(\boldsymbol{x},\boldsymbol{\xi},t)$ is the single-particle
distribution function, $\nabla f\equiv\frac{\partial}{\partial x_{\alpha}}$
is the spatial gradient of $f$, $\boldsymbol{\xi}$ is the microscale
velocity, $\tau$ is the relaxation time of the collision process,
and $f^{eq}=f^{eq}(\boldsymbol{x},\boldsymbol{\xi},t)$ is the local
Maxwell-Boltzmann (equilibrium) distribution function. Equation \ref{eq:BBGK}
is continuous in velocity and configuration $(\boldsymbol{x},t)$
space. To discretize the velocity space $\boldsymbol{\xi}$, the equation
is non-dimentionalized using a chosen speed of sound, and the resulting
$f^{eq}$ is expanded in a Taylor-series of fluid velocity $\boldsymbol{u}$
up to second-order. The discrete velocities are then obtained from
the requirement that the lower-order hydrodynamic moments with respect
to the truncated $f^{eq}$ satisfy the conservation of mass, momentum,
and energy \citep{He1997,Shan2006Kinetic}. Following this procedure,
we obtain the widely-used discrete velocity set of the\emph{ }D2Q9
lattice, for which the discrete Boltzmann-BGK equation can be written
as: 

\begin{equation}
\frac{\partial f_{i}}{\partial t}+\boldsymbol{\xi_{i}}\cdot\nabla f_{i}=-\frac{1}{\tau}(f_{i}-f_{i}^{eq}),\label{eq:DBE}
\end{equation}
where $f_{i}\equiv f_{i}(\boldsymbol{x},\boldsymbol{\xi}_{i},t)$,
$f_{i}^{eq}\equiv f_{i}^{eq}(\boldsymbol{x},\boldsymbol{\xi}_{i},t)$
and $i=0,1,2\cdots,8$. Here, while the Greek subscripts $\alpha\equiv\{x,y\}$
in 2D imply summation, the Latin subscripts (over velocity) \emph{do
not} imply summation. Equation \ref{eq:DBE} is termed as the \emph{discrete
Boltzmann equation} (DBE), and for a general class of discrete velocities,
also referred to as the discrete velocity model (DVM). The D2Q9 velocity
set is given by:

\begin{equation}
\boldsymbol{\xi}_{i}=\begin{cases}
(0,0) & \text{for\,}i=0\\
\sqrt{3}c_{s}\cos(\theta_{i},\sin\theta_{i}) & \text{for}\, i=1,2,3,4\\
\sqrt{3}c_{s}(\sqrt{2}(\cos\theta_{i},\sin\theta_{i})) & \,\text{for}\, i=5,6,7,8
\end{cases}
\end{equation}
where $\theta_{i}=(i-1)\pi/2$ for $i=1-4$, $\theta_{i}=(2i-9)\pi/4$
for $i=5-8$, and $c_{s}=1/\sqrt{3}$ is the speed of sound in the
lattice. Figure \ref{fig:D2Q9} shows a representation of the D2Q9
lattice.

The discrete form of the equilibrium distribution function (EDF) is
given by:

\begin{equation}
f_{i}^{eq}=\rho w_{i}\left[1+3\frac{\boldsymbol{\xi_{i}\cdot u}}{c^{2}}+\frac{9(\boldsymbol{\xi_{i}\cdot u})^{2}}{2c^{4}}-\frac{3\boldsymbol{u}^{2}}{2c^{2}}\right],\label{eq:Discrete EDF}
\end{equation}
where the weights, $w_{i}$, are:

\begin{equation}
w_{i}=\begin{cases}
4/9 & \text{for}\, i=0,\\
1/9 & \text{for}\, i=1,2,3,4,\\
1/36 & \text{\text{for}}\, i=5,6,7,8.
\end{cases}\label{eq:Weights}
\end{equation}
The macroscale density and velocity are related to the DF through:

\begin{eqnarray}
\rho & = & \sum_{i=0}^{8}f_{i}=\sum_{i=0}^{8}f_{i}^{eq}\\
\rho\boldsymbol{u} & = & \sum_{i=0}^{8}\boldsymbol{\xi}_{i}f_{i}=\sum_{i=0}^{8}\boldsymbol{\xi_{i}}f_{i}^{eq}.\nonumber 
\end{eqnarray}

It can be shown that a Chapman-Enskog expansion with the above discrete
form of the EDF recovers the incompressible, isothermal Navier-Stokes
equation in the limit of small Knudsen and Mach numbers with a shear
viscosity $\nu$ given by: 

\begin{equation}
\nu=c_{s}^{2}\tau
\end{equation}
where $\tau$ is the non-dimensional relaxation time.

\subsection{Off-Lattice Boltzmann Schemes}

\subsubsection{Explicit Runge-Kutta based schemes}

Since the discrete velocities $\boldsymbol{\xi}_{i}$ are constants,
the DBE can be considered as a system of linear, first-order, ordinary
differential equations (ODEs) with a weak source (collision) term.
This assumption is generally valid only if the gradients of the conserved
quantities in the flow are not too high, i.e., the collision term
is not highly non-linear. The number of ODEs in the system equals
the number of discrete velocities; for example; nine equations in
case of the \emph{D2Q9} lattice. Therefore, in principle, commonly
used time marching schemes for ODEs, such as Euler, Runge-Kutta, etc.
can be employed for temporal discretization of Equation \ref{eq:BBGK}.
On the other hand, FD or FV methods can be used for spatial discretization. 

Focusing on temporal discretization, a general second-order Runge-Kutta
(RK2) based OLB scheme for Equation \ref{eq:DBE} can be written as:

\begin{eqnarray}
f_{i}^{n+\frac{1}{2}} & = & f_{i}^{n}-\frac{\Delta t}{2}R_{i}^{n}\label{eq:RK2}\\
f_{i}^{n+1} & = & f_{i}^{n}-\Delta tR_{i}^{n+\frac{1}{2}}\nonumber 
\end{eqnarray}
where

\begin{equation}
R_{i}^{n}\equiv-\left(\boldsymbol{\xi}_{i\alpha}\frac{\partial f_{i}^{n}}{\partial x_{\alpha}}\right)-\frac{1}{\tau}(f_{i}^{n}-f_{i}^{eq,n})
\end{equation}
with the gradient term expanded in 2D Cartesian co-ordinates as: 

\begin{equation}
\boldsymbol{\xi}_{i}\cdot\nabla f_{i}^{n}=\boldsymbol{\xi}_{i\alpha}\frac{\partial f_{i}^{n}}{\partial x_{\alpha}}=\xi_{ix}\frac{\partial f_{i}^{n}}{\partial x}+\xi_{iy}\frac{\partial f_{i}^{n}}{\partial y}.
\end{equation}

Here $f_{i}^{n+\frac{1}{2}}\equiv f_{i}(\boldsymbol{x},t_{n}+\frac{\Delta t}{2})$,
$f_{i}^{n+1}\equiv f_{i}(\boldsymbol{x},t_{n}+\Delta t)$, etc. Similar
expressions can be written for the fourth-order RK scheme (RK4) \citep{Reider95accuracyof}.
In these schemes, the viscosity was related to the relaxation time
through $\nu=c_{s}^{2}\tau$. 

Many of the earliest OLB schemes employed the forward Euler, RK2 or
RK4 schemes, in combination with a variety of spatial discretization
schemes. Using the RK-based time marching schemes, the geometric flexibility
of the LB method was extended to non-Cartesian domains, non-uniform
grids, body-fitted, stretched grids, etc. \citep{Cao1997Physical,Reider95accuracyof,Kandhai2001,So2010Finite,Tamura20113D}.
In the case of FD spatial discretization, the spatial order-of-accuracy
can also be increased arbitrarily, using higher-order Taylor approximations
of the gradient terms. On the other hand, several discrete velocity
models with non-space-filling velocity-sets also employed the RK-based
schemes as the evolution equation \citep{Watari2004}.

Despite the geometric flexibility made possible by the RK-based schemes,
the size of the $\Delta t$ relative to $\tau$ has to be kept very
small to maintain numerical stability. The constraint on $\Delta t$
comes primarily from the \emph{explicit} approximation of the advection
(LHS) and collision (RHS) terms of the DBE. An explicit advection
approximation imposes a stability criterion on the size of $\Delta t$
through the \emph{CFL} condition, $CFL=\boldsymbol{\xi}_{i}\Delta t/\Delta x<1$.
The $CFL$ condition is well-understood to be a necessary condition
for the stability of advection type of equation. However, the overall
stability of the scheme is governed by the more restrictive condition
on $\Delta t$ due to explicit approximation of collision, given by
the approximate condition $\Delta t<\tau$ \citep{Ubertini2008Generalised,Lee2003Eulerian}.

\subsubsection{Characteristics-based schemes}

The characteristics-based OLB schemes are based on time integration
of the DBE along the \emph{characteristics} using the $\theta-$method
\citep{Bardow2006General,Lee2003Eulerian}:

\begin{equation}
\tilde{f}_{i}^{n+1}=\tilde{f}_{i}^{n}+\Delta t\left[\left(1-\theta)\tilde{\Omega}_{i}^{n}\right)+\theta\tilde{\Omega}_{i}^{n+1}\right],\label{eq:theta method}
\end{equation}
where we denote $\Omega_{i}=\frac{1}{\tau}(f_{i}-f_{i}^{eq})$ for
brevity; a tilde indicates a term on the characteristic line, i.e.,
$\tilde{f}_{i}^{n+1}=f_{i}(\boldsymbol{x}+\boldsymbol{\xi}_{i}\Delta t,t+\Delta t)$,
$\tilde{f}_{i}^{n}=f_{i}(\boldsymbol{x},t)$; and $0\leq\theta\leq1$.
For $\theta=\{0,\frac{1}{2},1\}$, we obtain an explicit $\mathcal{O}(\Delta t)$,
an implicit $\mathcal{O}(\Delta t^{2})$, and an implicit $\mathcal{O}(\Delta t)$
approximation of the collision term, respectively. The $\mathcal{O}(\Delta t)$
schemes are Euler-type schemes, and the $\mathcal{O}(\Delta t^{2})$
is a Crank-Nicholson-type scheme. However, a $\mathcal{O}(\Delta t^{2})$
collision approximation can still be obtained for any value of $\theta\in[0,1]$,
if the viscosity and relaxation time are related through: 

\begin{equation}
\nu=\left(\frac{\tau}{\Delta t}-0.5+\theta\right)c_{s}^{2}\Delta t.\label{eq:visc-modrelaxtime}
\end{equation}
A distinction should be made on the order of magnitude of $\tau$
as used in the standard LBM versus in the OLB method. In the standard
LBM, $\tau$ is $\mathcal{O}(1)$, and in fact typically in the range
of $0.6<\tau<3.5$. In the OLB method, on the other hand, due to non-dimentionalization,
$\tau$ is $\mathcal{O}(Kn)$, where \emph{Kn} is the Knudsen number.
Therefore, in OLB method, $\tau$ is $\mathcal{O}(0.01)$ or lower,
depending upon the \emph{Re.} 

Broadly based on Equation \ref{eq:theta method}, several OLB schemes
have been proposed that are numerical stable at much larger $\Delta t/\tau$
. In general, these schemes differ in their advection and collision
approximations (explicit or implicit), as described below.

Following Equation \ref{eq:theta method}, Lee and Lin \citep{Lee2003Eulerian}
proposed a \emph{fully-explicit} scheme (advection and collision),
termed herein as AE/CE, which can be written as: 

\begin{eqnarray}
f_{i}^{n+1} & = & f_{i}^{n}-\Delta t\left[\xi_{i\alpha}\frac{\partial f_{i}^{n}}{\partial x_{\alpha}}+\frac{1}{\lambda}(f_{i}^{n}-f_{i}^{eq,n})\right]+\Delta t^{2}\xi_{i\alpha}\frac{\partial}{\partial x_{\alpha}}\left[\frac{1}{2}\xi_{i\beta}\frac{\partial f_{i}^{n}}{\partial x_{\beta}}+\frac{1}{\lambda}(f_{i}^{n}-f_{i}^{eq,n})\right]\\
 &  & -\frac{\Delta t^{3}}{2}\frac{1}{\lambda}\xi_{i\alpha}\frac{\partial}{\partial x_{\alpha}}\left[\xi_{i\beta}\frac{\partial}{\partial x_{\beta}}(f_{i}^{n}-f_{i}^{eq,n})\right],\nonumber 
\end{eqnarray}
with $\lambda=\nu/c_{s}^{2}+0.5\Delta t$ corresponding to $\theta=0$,
where $\lambda$ is also termed as the modified-relaxation parameter. 

It is well-known in the ODE theory that an implicit approximation
of the relaxation (collision) term is critical for numerical stability,
especially for \emph{stiff} equations (highly non-linear flows). Following
this fact, a scheme similar to AE/CE but with an \emph{implicit} collision
(CI) treatment i.e., $\theta=1/2$ was proposed in \citep{Lee2001Characteristic},
herein referred to as the AE/CI scheme. The AE/CI scheme, however,
required an iterative predictor-corrector type of approximation for
collision, which increases the computational cost and complexity of
the simulations. 

In order to avoid an iterative procedure to approximate an implicit-collision,
a variable transformation is often employed in the LB method to mask
the implicitness of the collision term \citep{He1998,Bosch2013}.
A similar technique is employed in the OLB context, where a new DF,
$g_{i}$, is defined as: 

\begin{equation}
g_{i}(\boldsymbol{x},t)=f_{i}(\boldsymbol{x},t)-\Delta t\theta\Omega_{i}(f_{i}(\boldsymbol{x},t)).\label{eq:variable transf.}
\end{equation}
Importantly, the variable transformation process preserves mass and
momentum conservation, i.e., $\rho=\sum_{i}f_{i}=\sum_{i}g_{i}$ and
$\rho\boldsymbol{u}=\sum_{i}\boldsymbol{\xi}_{i}f_{i}=\sum_{i}\boldsymbol{\xi}_{i}g_{i}$.
The variable transformation with $\theta=\frac{1}{2}$ also maintains
$\mathcal{O}(\Delta t^{2})$ accuracy of the collision approximation
\citep{Ubertini2010Three}.

With the variable transformation transformation technique and $\theta=\frac{1}{2}$,
Guo and Zhao \citep{Guo2003Explicit} proposed an OLB scheme that
can be written as:

\begin{equation}
g_{i}^{n+1}=f_{i}^{n}-\Delta t\xi_{i\alpha}\frac{\partial f_{i}^{n}}{\partial x_{\alpha}}-\frac{\Delta t}{2\tau}(f_{i}^{n}-f_{i}^{eq,n}).\label{eq:Guo Scheme}
\end{equation}
In this scheme, although the collision is \emph{implicit} and $\mathcal{O}(\Delta t^{2})$,
the advection is \emph{explicit} and $\mathcal{O}(\Delta t)$ along
the characteristics. Following Equation \ref{eq:visc-modrelaxtime}
for $\theta=1/2$, the viscosity-relaxation time relation in this
scheme is $\nu=\tau c_{s}^{2}$. In this work, this scheme is referred
as GZ scheme.

Bardow \emph{et al.} \citep{Bardow2006General} combined the variable
transformation technique, along with an explicit $\mathcal{O}(\Delta t^{2})$
advection approximation along the characteristics to yield the BKG
scheme:

\begin{eqnarray}
g_{i}^{n+1} & = & g_{i}^{n}-\Delta t\left[\xi_{i\alpha}\frac{\partial g_{i}^{n}}{\partial x_{\alpha}}+\frac{1}{\lambda}(g_{i}^{n}-g_{i}^{eq,n})\right]+\frac{\Delta t^{2}}{2}\xi_{i\alpha}\frac{\partial}{\partial x_{\alpha}}\left[\xi_{i\beta}\frac{\partial g_{i}^{n}}{\partial x_{\beta}}+\frac{2}{\lambda}(g_{i}^{n}-g_{i}^{eq,n})\right]\label{eq:Bardow Scheme}\\
 & - & \frac{\Delta t^{3}}{2\lambda}\xi_{i\alpha}\frac{\partial}{\partial x_{\alpha}}\left[\xi_{i\beta}\frac{\partial(g_{i}^{n}-g_{i}^{eq,n})}{\partial x_{\beta}}\right].\nonumber 
\end{eqnarray}
Due to the implicit treatment of the collision term through variable
transformation, the $\Delta t<\tau$ constraint no longer applies.
The only constraint on $\Delta t$ is the \emph{CFL} condition due
to explicit treatment of advection. The viscosity is related to the
relaxation time through $\lambda=\nu/c_{s}^{2}+0.5\Delta t$. A similar
scheme called the unstructured lattice Boltzmann with memory (ULBEM)
was proposed by \citep{Ubertini2008Generalised}. The various schemes
described above are summarized in Table 1.

\begin{center}
\begin{table}[H]
\caption{Summary of various off-lattice Boltzmann schemes. Here \emph{E} indicates
explicit, and \emph{I }indicates implicit. Spatial order-of-accuracy
for advection depends upon the spatial discretization method and schemes
used, and hence is excluded. }

\centering{}%
\begin{tabular}{|c|c|c|c|}
\hline 
Scheme & Advection & Collision & $\nu-\lambda$ relation\tabularnewline
\hline 
\hline 
RK2/RK4 & $\mathcal{O}(\Delta t^{2})$ E & $\mathcal{O}(\Delta t^{2})$ E & $\lambda=\nu/c_{s}^{2}$\tabularnewline
\hline 
AE/CE  & $\mathcal{O}(\Delta t^{2})$ E & $\mathcal{O}(\Delta t^{2})$ E & $\lambda=\nu/c_{s}^{2}+0.5\Delta t$\tabularnewline
\hline 
AE/CI & $\mathcal{O}(\Delta t^{2})$ E & $\mathcal{O}(\Delta t^{2})$ I & $\lambda=\nu/c_{s}^{2}$\tabularnewline
\hline 
GZ & $\mathcal{O}(\Delta t)$ E & $\mathcal{O}(\Delta t^{2})$ I & $\lambda=\nu/c_{s}^{2}$\tabularnewline
\hline 
BKG & $\mathcal{O}(\Delta t^{2})$ E & $\mathcal{O}(\Delta t^{2})$ I & $\lambda=\nu/c_{s}^{2}+0.5\Delta t$\tabularnewline
\hline 
\end{tabular}
\end{table}

\par\end{center}

\subsection{Implementation aspects of a OLB scheme}

Equation \ref{eq:Bardow Scheme} represents a typical temporal evolution
scheme for the DFs used in the OLB method. Clearly, the usual stream-collide
scheme type of evolution on the standard LBM is replaced by a procedure
that involves solving a set of ODEs at each node. For the BKG scheme,
the algorithm consists of the following steps:
\begin{enumerate}
\item Initialize $f_{i}^{0}$ and $g_{i}^{0}$ according to prescribed initial
conditions, i.e., set $f_{i}^{0}=g_{i}^{0}=g_{i}^{eq,0}$, where $g_{i}^{eq,0}=g_{i}^{eq}(\rho^{0},\boldsymbol{u}^{0})$.
$\rho^{0}$ and $\boldsymbol{u}^{0}$ are the given initial conditions.
\item For a particular time-step $n$, evaluate $g_{i}^{n}$ through Equation
\ref{eq:variable transf.} (variable transformation).
\item Evaluate the gradient terms $\frac{\partial g_{i}^{n}}{\partial x},$
$\frac{\partial^{2}g_{i}^{n}}{\partial x^{2}}$, etc. In the case
of FD spatial discretization, with a central-differencing scheme,
these quantities can be computed as:

\begin{eqnarray}
\frac{\partial g_{i}^{n}}{\partial x} & = & \frac{g_{i}(x+\Delta x,y,t_{n})-g_{i}(x-\Delta x,y,t_{n})}{2\Delta x}+\mathcal{O}(\Delta x^{2}),\nonumber \\
\frac{\partial^{2}g_{i}^{n}}{\partial x^{2}} & = & \frac{g_{i}(x+\Delta x,y,t_{n})-2g_{i}(x,y,t_{n})+g_{i}(x-\Delta x,y,t_{n})}{\Delta x^{2}}+\mathcal{O}(\Delta x^{2}).\label{eq:Spatial gradients}
\end{eqnarray}
Similar expressions can be written for the first-order derivatives
in the \emph{y}-direction, and the second-order mixed derivatives
$\frac{\partial^{2}g_{i}}{\partial x\partial y}$ \citep{Hirsch2007}.
Other schemes such as upwind-differencing can also be employed. 

\item Evaluate $g_{i}^{n+1}$ per Equation \ref{eq:Bardow Scheme}. 
\item Evaluate $\rho^{n+1}$, $\boldsymbol{u}^{n+1}$ and $g_{i}^{eq,n+1}$. 
\item Evaluate $f_{i}^{n+1}$ per equation \ref{eq:variable transf.}.
\item If $\boldsymbol{u}^{n+1}$ has converged, then stop, if not, repeat
steps 2-6.
\end{enumerate}

\section{Numerical Tests}

To test the numerical stability of the various schemes, as a function
of the maximum stable $\Delta t/\tau$, several steady and unsteady,
incompressible, two-dimensional flows were simulated, and their results
are presented in this section. Since the focus of this work is on
evaluation of the temporal schemes, the simpler finite-difference
method is used to discretize the spatial domains. For uniformity,
Equation \ref{eq:Spatial gradients} is used to evaluate the derivatives
in the gradient term for all the OLB schemes tested here. The central-difference
scheme is chosen since they are less diffusive than other $\mathcal{O}(\Delta x^{2})$
schemes. This minimizes the contribution of numerical diffusion to
the overall stability of the scheme. The numerical stability of a
scheme can be concluded by the maximum allowable $\Delta t$ for a
particular $Re$ and grid size. In other words, for a particular flow
problem, we fix the $Re$ (fixed $\tau$) and grid size (fixed $\Delta x)$
and vary $\Delta t$, until the simulation becomes unstable. 

The schemes were coded in C++ using the Armadillo linear algebra library
\citep{Armadillo}, and parallelized for a shared-memory architecture
using OpenMP.

\subsection{Taylor Vortex Flow}

To test the stability and accuracy of various schemes without the
artifacts of a boundary treatment, we first simulate the Taylor-Green
vortex flow. This flow represents the unsteady flow of a freely decaying
two-dimensional vortex, and is often used to evaluate the effective
viscosity and temporal and spatial accuracy of a scheme. Here, the
flow is computed within a periodic square box defined as $-\pi\leq x,y\leq\pi$
with a uniform Cartesian mesh of size $100\times100$ on the domain.
The analytical solutions of the flow-field are given by:

\begin{eqnarray}
u(x,y,t) & = & -u_{ref}\cos(k_{1}x)\sin(k_{2}y)\exp[-\nu(k_{1}^{2}+k_{2}^{2})t],\\
v(x,y,t) & = & u_{ref}\frac{k_{1}}{k_{2}}\sin(k_{1}x)\cos(k_{2}y)\exp[-\nu(k_{1}^{2}+k_{2}^{2})t],\nonumber \\
p(x,y,t) & = & p_{ref}-\frac{u_{0}^{2}}{4}\left[\cos(2k_{1}x)+\frac{k_{1}^{2}}{k_{2}^{2}}\cos(2k_{2}y)\right]\exp[-2\nu(k_{1}^{2}+k_{2}^{2})t].\nonumber 
\end{eqnarray}

In our simulations, to minimize compressibility effects, the Mach
number is set as $0.01$. The reference velocity is $u_{ref}=Ma\times c_{s}$,
the reference density is $\rho_{ref}=1$, the reference pressure is
$p_{ref}=\rho_{ref}/c_{s}^{2}$, and the wave numbers $k_{1}$ and
$k_{2}$ are constants set to $1.0$ and $4.0$, respectively. The
non-equilibrium initialization scheme proposed by \citep{Skordos1993}
is used to initialize the DFs, and periodic boundary conditions are
applied in the $x$ and $y$ directions. The \emph{$Re$} of the flow
is $100$. 

Figures \ref{fig:TVF_ux} and \ref{fig:TVF_uy} show the horizontal
and vertical velocity profiles at different times as obtained using
the BKG scheme. The velocity profiles are shown for $t/t_{c}=0.5,1,$
and $2$, where $t_{c}=\ln2/\nu(k_{1}^{2}+k_{2}^{2})$ is the time
at which the amplitude of the decay is halved. Importantly, these
simulation have been obtained with $\Delta t=10\tau$, where $\tau=\nu/c_{s}^{2}$.
The corresponding $CFL$ number is $0.7$. This confirms the observations
of high $\Delta t/\tau$ made in \citep{Bardow2006General}. Moreover,
the average relative error as defined by: 

\[
E=\frac{\sum_{y}\left(|u_{num}-u_{exact}|+|v_{num}-v_{exact}|\right)}{\sum_{,y}\left(|u_{exact}|+|v_{exact}|\right)}
\]
is $<1\%$. Here $u_{num}$ is the simulated velocity, $u_{exact}$
is the analytical velocity, and the summation is over vertical plane
at $x=0$. This result demonstrates that the BKG scheme successfully
overcomes the $\Delta t<\tau$ restriction imposed by the RK-based
OLB schemes. The only remaining restriction on $\Delta t$ is the
\emph{local} $CFL$ number which is due to an explicit advection approximation. 

\begin{center}
\begin{figure}[H]
\begin{minipage}[c][1\totalheight][t]{0.45\textwidth}%
\begin{center}
\includegraphics[scale=0.4]{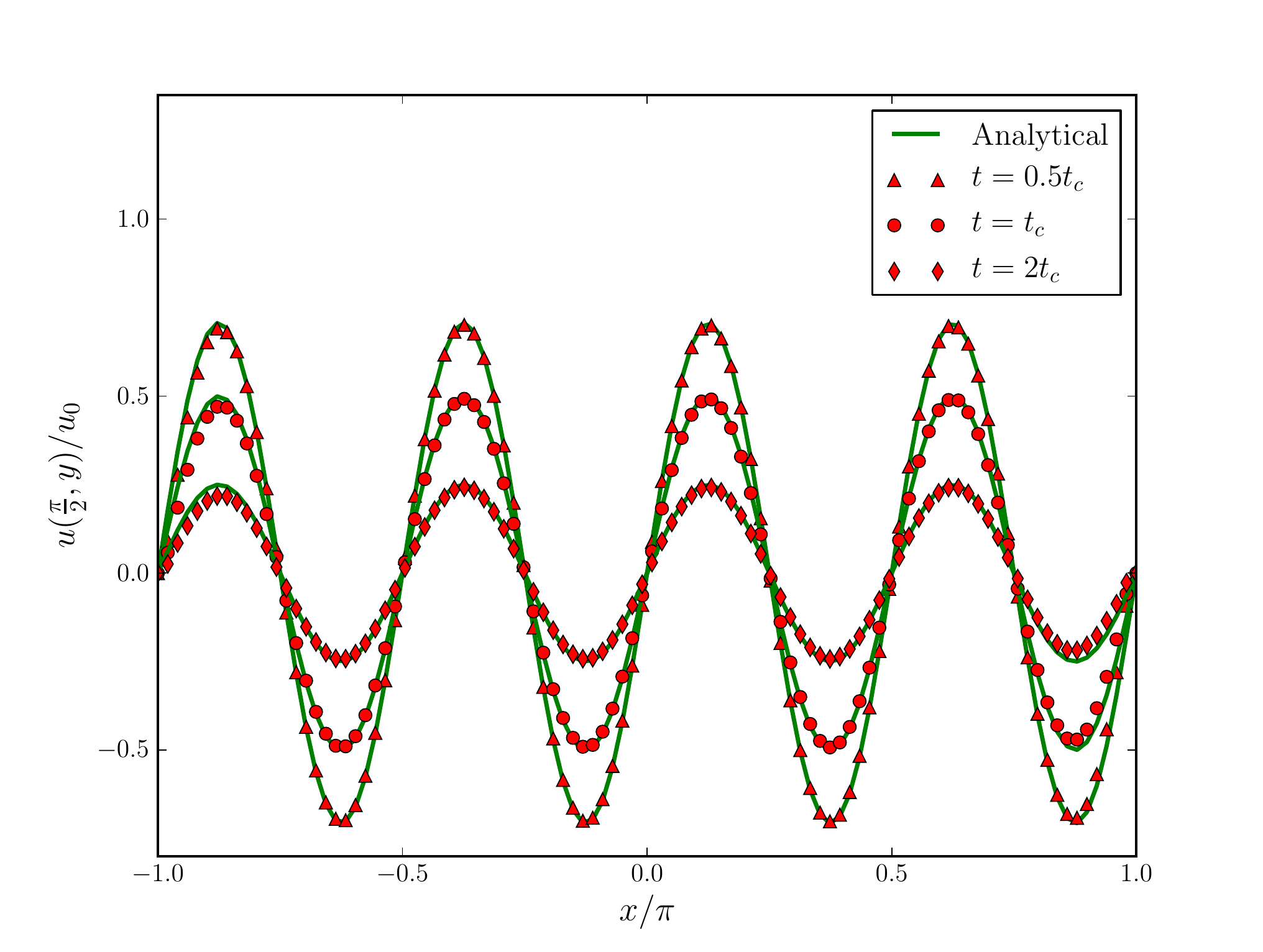}
\par\end{center}

\caption{Horizontal velocity profile of Taylor-Vortex flow at various times.
The numerical results have been obtained with a $\Delta t=10\tau$
using the BKG scheme.\label{fig:TVF_ux}}
\end{minipage}\hfill{}%
\begin{minipage}[c][1\totalheight][t]{0.45\textwidth}%
\begin{center}
\includegraphics[scale=0.4]{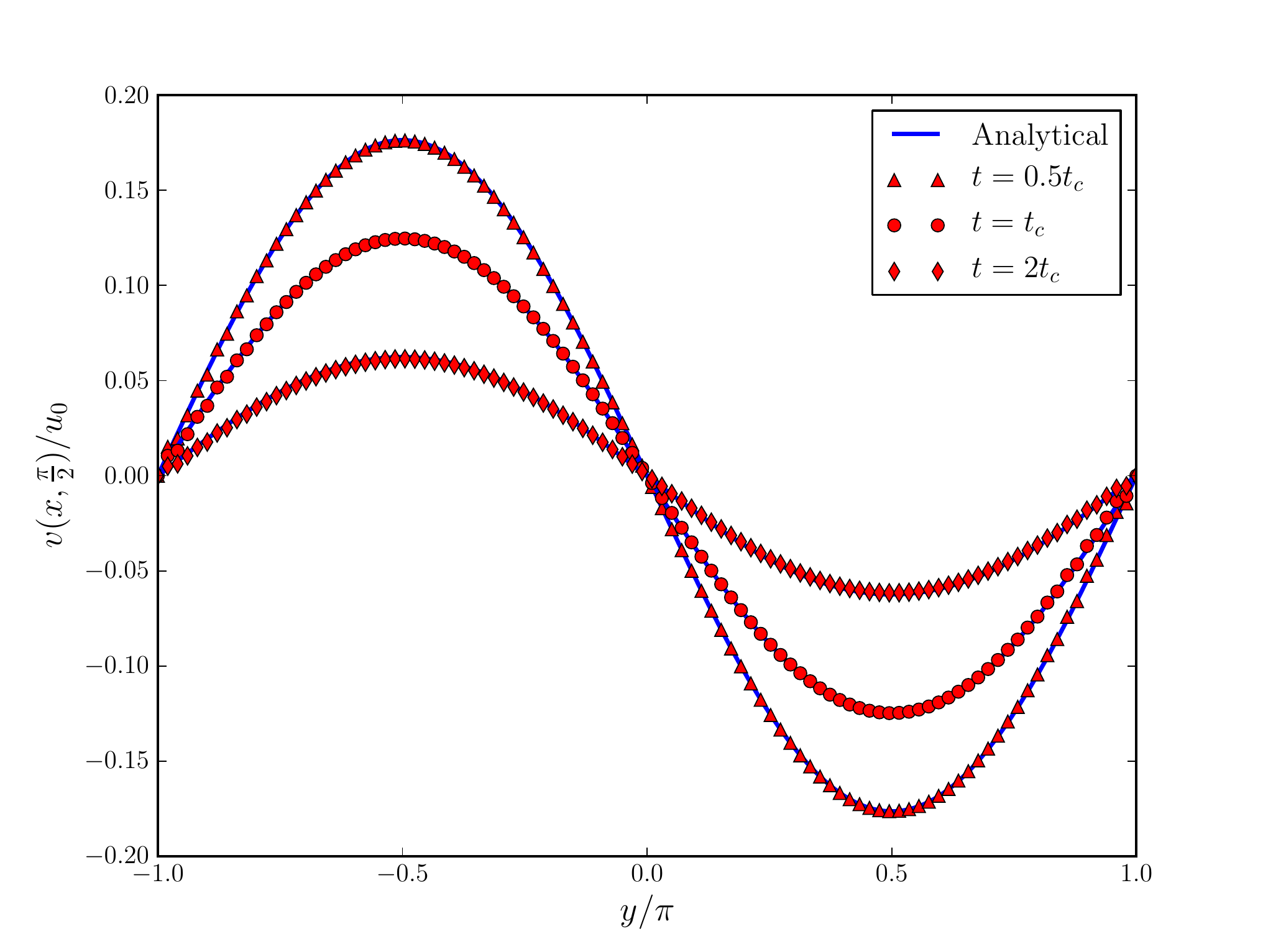}
\par\end{center}

\caption{Vertical velocity profile of Taylor-Vortex flow at various times with
$\Delta t=10\tau$ using the BKG scheme.\label{fig:TVF_uy}}
\end{minipage}
\end{figure}

\par\end{center}

For comparison, we also simulated the Taylor-Green vortex flow using
the GZ and the AE/CE schemes. The GZ scheme, although having treated
the collision term implicitly, could not achieve higher $\Delta t/\tau$
values. In fact, in their simulations, $\Delta t/\tau\approx1.6$
for a $Re\approx63$ with a mixed (central-upwind) differencing scheme.
Furthermore, for a purely central-difference scheme, i.e., without
the additional marginal stability of upwind schemes, the authors showed
analytically using the von-Neumann stability analysis, that the maximum
$\Delta t/\tau$ that can be attained is less than $1.5$, and that
at low \emph{$CFL$} numbers. We have observed that the AE/CE permits
$\Delta t>\tau$, but stable simulations are obtained at much lower
values of $CFL$ number ($CFL<0.2)$ compared to the BKG AE/CI scheme.

\subsection{2-D Plane Poiseuille Flow}

Another useful numerical test is Poiseuille flow. A plane Poiseuille
flow describes the steady, laminar flow of an incompressible fluid
in a rectangular channel driven by a pressure gradient. Assuming symmetry
and incompressibility, it can be shown that for Poiseuille flow, the
Navier-Stokes momentum equation reduces to:

\begin{equation}
\mu\frac{\partial^{2}u}{\partial y^{2}}=\frac{\partial p}{\partial x}
\end{equation}
which has a exact steady state solution for velocity given by:

\begin{eqnarray}
u(y) & = & 4u_{max}\frac{y}{H}\left(1-\frac{y}{H}\right)\,\,\,\,\,\text{for }0\leq y\leq H\label{eq:Poisuille vel eq}\\
v(x,y) & = & 0\nonumber 
\end{eqnarray}
where, $H$ is the channel height and $u_{max}$ is the center-line
velocity where the magnitude of velocity is the maximum. The Reynolds
number of the flow is defined by $Re=u_{max}H/\nu$. In a LB simulation,
imposing a pressure difference by specifying inlet and outlet densities
increases the compressibility errors \citep{Succi2001}. Therefore,
the effect of $\Delta p$ is imposed on the flow through an equivalent
body-force $F$. This force has the same effect as having a pressure-gradient
in the channel which produces the chosen $u_{max}$. This body force
can be evaluated from $F=8u_{max}\rho\nu/H^{2}$. 

For the reference case, the BKG scheme is used to obtain the results
presented here. In the simulation, we set $H=1$ and $L=1$, where
$L$ is the channel length. The domain is discretized into a uniform
Cartesian mesh of size $100\times100$. The Mach number based on $u_{max}$
is set to $0.1732$, which corresponds to a $u_{max}$ of $0.1$.
The velocity is initialized to zero everywhere and the average density
is set to one. Since a steady-state flow is simulated, the equilibrium
initial condition is applied to the DFs. A non-equilibrium extrapolation
boundary scheme is applied on the no-slip top and bottom walls \citep{Guo2007ExtrapolationBC},
and a periodic boundary condition is applied at the inlet and outlets.
Various $Re$ flows are simulated by varying the viscosity but keeping
the $Ma$ constant. The simulations are run until a steady state criterion,
defined as:

\[
C=\frac{\sum_{_{i,j}}|u_{i,j}^{n}-u_{i,j}^{n-50}|}{\sum_{i,j}|u_{i,j}^{n}|}<10^{-6}
\]
is attained. Here $u_{i,j}^{n}\equiv u(x_{i},y_{j},n\Delta t)$, where
$n$ is the time-step number, and the summation is over the entire
flow field.

The simulated steady-state stream-wise velocity profile for $Re=100$,
along with the analytical solution is shown in Figure \ref{fig:Pois_ux_vs_y}.
Clearly, the simulated and the analytical values are in excellent
agreement with each other. The inset shows the velocity profile close
to the wall, which is also in close agreement to analytical values.
This serves as validation for the applicability and accuracy of the
non-equilibrium boundary condition used in the simulation. It is worth
noting that the velocity profile was generated on a $100\times100$
grid with a $\Delta t=5\tau$. Even at such a high $\Delta t$, the
average relative error as defined earlier is $\sim2\%$. 

\begin{center}
\begin{figure}[H]
\begin{centering}
\includegraphics[scale=0.6]{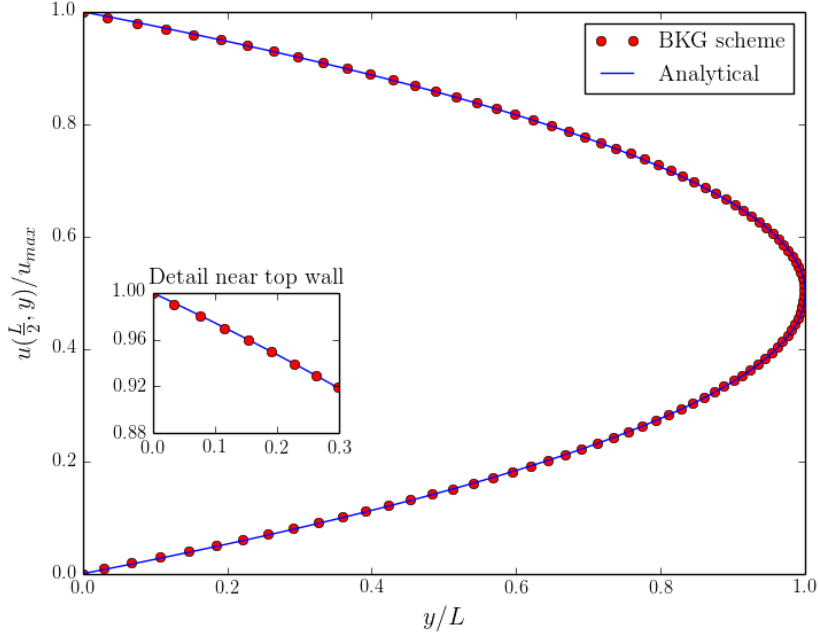}
\par\end{centering}

\caption{Steady-state horizontal velocity profile for Poiseuille flow at $Re=100$.
The results were obtained using $\Delta t=5\tau$ on a $100\times100$
mesh using the BKG scheme.\label{fig:Pois_ux_vs_y}}
\end{figure}

\par\end{center}

For comparison, the plane Poiseuille flow is also simulated using
the AE/CE scheme, and the GZ scheme. Figure \ref{fig:Maximum-allowable-CFL_Poiseulle}
shows the maximum allowable $CFL$ number versus $Re$ for the steady
Poiseuille flow for the different schemes. This plot is obtained by
fixing the grid size to $100\times100$ for each scheme, and then
varying $\Delta t$. This is repeated for the various $Re$, as shown
in the figure. 

It can be seen that at low $Re$, both the AE/CE and BKG schemes have
almost similar maximum $CFL$ numbers. However, as $Re$ increases,
the maximum allowable $CFL$ number is clearly much higher with the
BKG scheme, as compared to the AE/CE scheme. This enhanced stability
at higher \emph{$Re$} results from the implicit treatment of the
collision term, which becomes highly non-linear as $Re$ increases,
thus directly benefiting from the local implicit treatment. Importantly,
the implicit treatment of collision, allows for $\Delta t>\tau$,
thus resulting in large $\Delta t$ and thereby reducing the computational
effort. 

\begin{center}
\begin{figure}[H]
\begin{centering}
\includegraphics[scale=0.5]{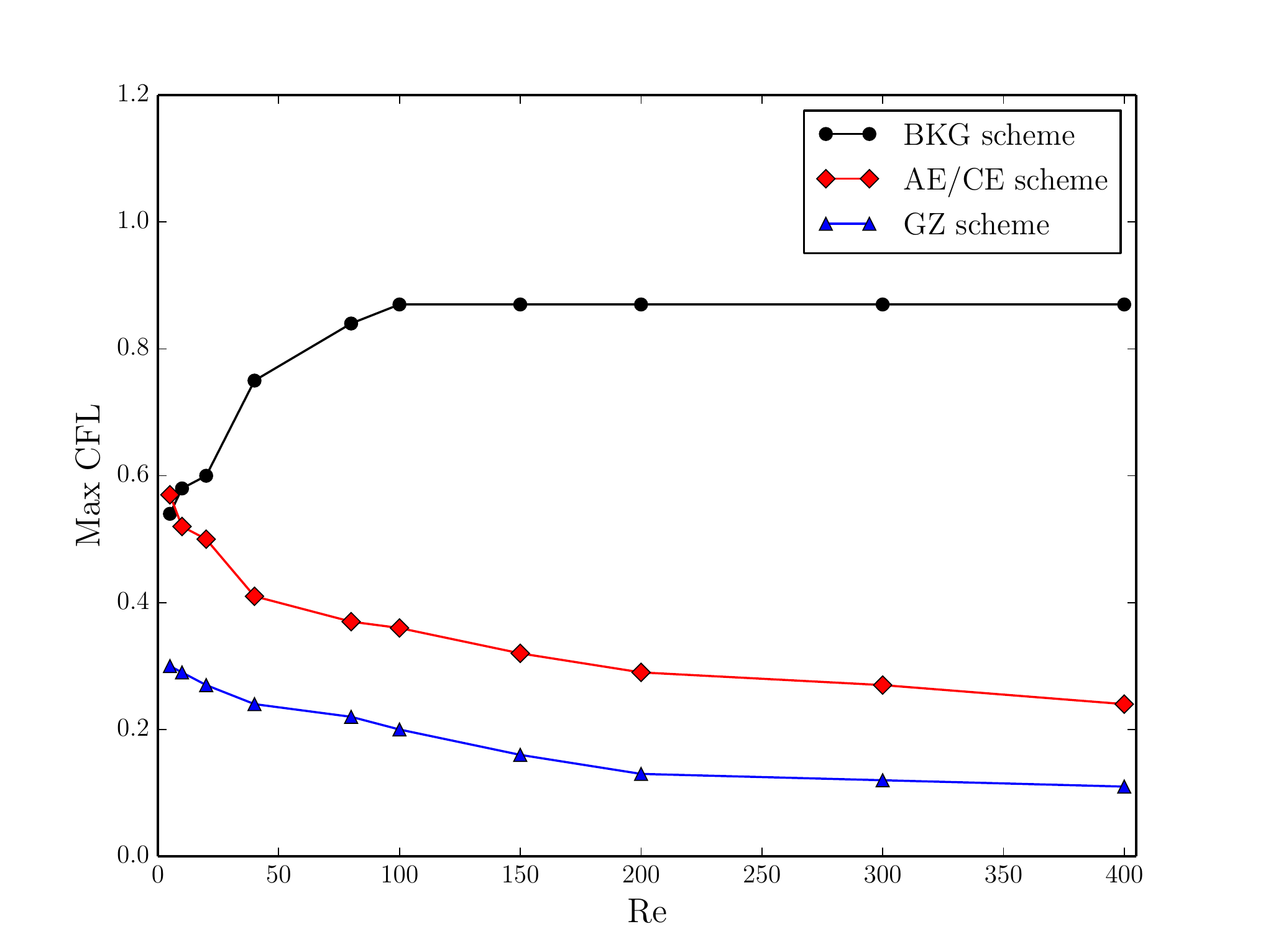}
\par\end{centering}

\caption{Maximum allowable $CFL$ number vs. \emph{Re} for steady Poiseuille
flow.\label{fig:Maximum-allowable-CFL_Poiseulle}}
\end{figure}

\par\end{center}

\subsection{Lid-Driven Cavity Flow}

Isothermal, 2-D lid-driven cavity flow is commonly used as a benchmark
case to test and evaluate numerical schemes for incompressible viscous
flows. The flow domain consists of a square cavity with three stationary
walls on the sides and bottom, and a top wall (lid) that moves with
a uniform tangential velocity $u_{lid}$. Despite the simple geometry,
this flow exhibits many complex flow patterns such as formation of
vortices near corners due to singularities. In our context, the lid-driven
cavity case is also useful for quantitatively evaluating the effects
of collision approximations (explicit/implicit) on flows with moderately
large non-linearities (high $Re$), and thus quantifying the overall
stability of different schemes. 

The computational domain consists of a square with height $L=H=1$.
The top wall moves with a constant velocity of $u_{lid}=0.1$, and
the Reynolds number of the flow is $Re=u_{lid}H/\nu$. The domain
is discretized on a uniform mesh of size $257\times257$. The non-equilibrium
extrapolation scheme is applied for all of the walls. The flow is
initialized by setting $\rho=1$ and $\boldsymbol{u}=0$ in the entire
flow. The flow is simulated using four schemes: BKG, AE/CE, GZ, and
also a RK2-based scheme.

As before, for the reference case, we present the results of simulations
using the BKG scheme. Steady-state velocity components along the horizontal
and vertical centerlines at various $Re$ are shown in Figures \ref{fig:LDC x-velocity}
and \ref{fig:LDC y-velocity}. These results have been obtained with
$\Delta t=4\tau$ and $CFL=0.5$. The results are compared with results
from Ghia \emph{et al.} who obtained their results with a $257\times257$
grid using the coupled strongly implicit multi-grid method, and a
vorticity-stream-function formulation \citep{Ghia1982} . The velocity
profiles change from curved at lower \emph{$Re$} to linear for higher
$Re$, which is consistent with the benchmark solutions. The near-linear
velocity profiles at higher \emph{$Re$} in the central core of the
cavity indicates a region of uniform vorticity. 

\begin{center}
\begin{figure}[H]
\begin{minipage}[c][1\totalheight][t]{0.45\textwidth}%
\begin{center}
\includegraphics[scale=0.42]{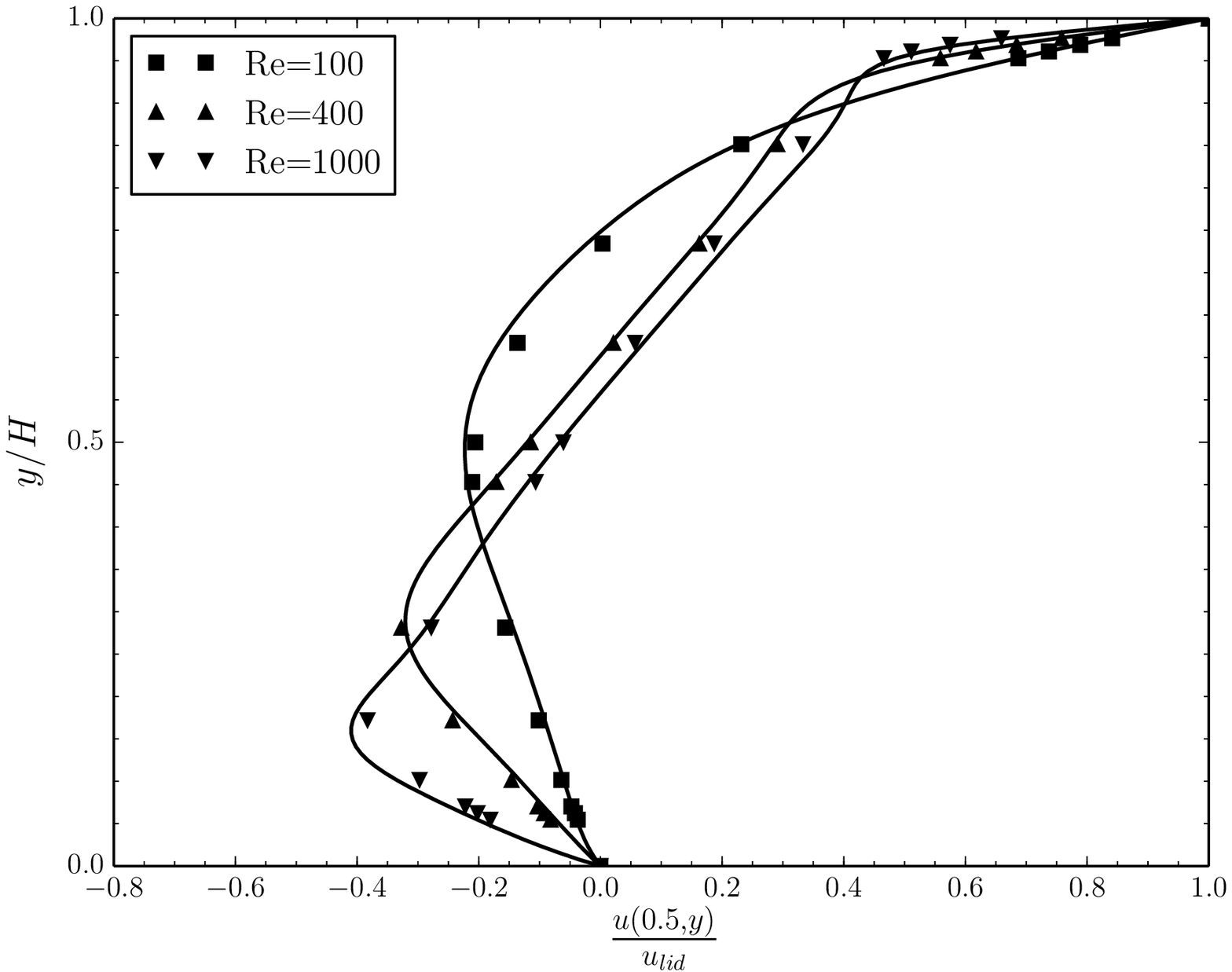}
\par\end{center}

\caption{Centerline horizontal-velocity profile for a lid-driven cavity at
various $Re$. The solid lines indicate the simulated values, while
the markers $(\blacksquare,\blacktriangle,\blacktriangledown)$ indicate
the corresponding solution from Ghia et al. \citep{Ghia1982}\label{fig:LDC x-velocity}}
\end{minipage}\hfill{}%
\begin{minipage}[c][1\totalheight][t]{0.45\textwidth}%
\begin{center}
\includegraphics[scale=0.42]{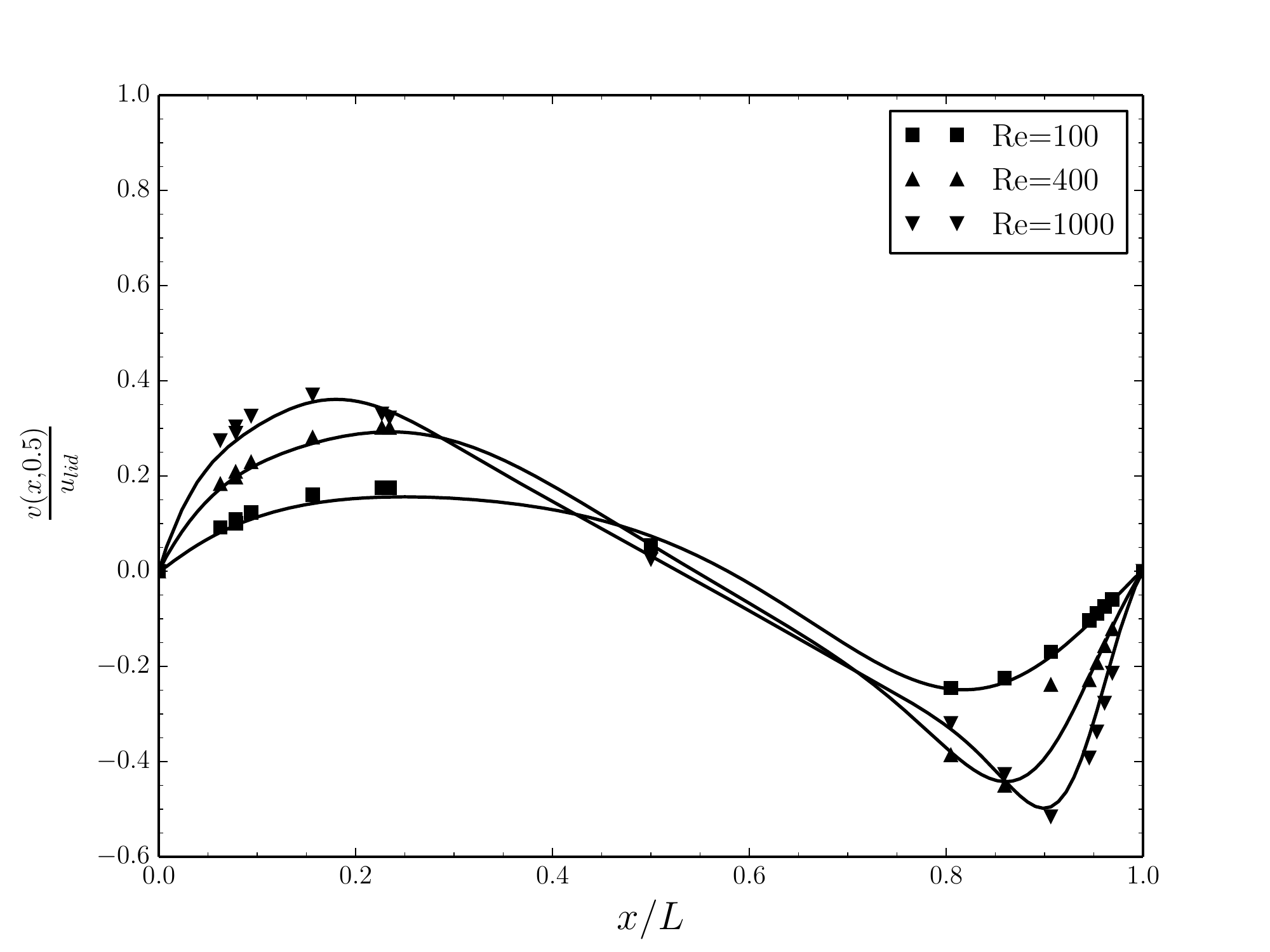}
\par\end{center}

\caption{Centerline vertical-velocity profile for a lid-driven cavity at various
$Re$. The solid lines indicate the simulated values, while the markers
$(\blacksquare,\blacktriangle,\blacktriangledown)$ indicate the corresponding
solution from Ghia et al. \citep{Ghia1982}\label{fig:LDC y-velocity}}
\end{minipage}
\end{figure}
 
\par\end{center}

To assess and quantify the effects of the size of $\Delta t$ on numerical
stability, in Figures \ref{fig:Stab_RK2}-\ref{fig:Stab_Bard_1000},
we plot the stability region for each scheme, as determined by two
parameters: $\Delta t/\tau$ and $\Delta t/\Delta x$ (please note
that the scales of the axes vary across the figures). Broadly speaking,
while $\Delta t/\tau$ represents the non-linear stability constraint
imposed by collision, $\Delta t/\Delta x$ represent the linear-stability
constraints of explicit advection. $\Delta t/\Delta x$ also represents
the $CFL$ number in the case of a D2Q9 lattice. Hence, a map determined
by these two parameters should serve a guide to gauge the overall
numerical stability of the schemes with respect to size of $\Delta t$.
In all of the maps, $\bullet$ indicates a stable solution and $\circ$
indicates an unstable solution. Figure \ref{fig:Stab_RK2} shows the
stability region for the RK2-based OLB scheme with central differencing
for $Re=100$. 

\begin{center}
\begin{figure}[H]
\begin{minipage}[c][1\totalheight][t]{0.45\textwidth}%
\begin{center}
\includegraphics[scale=0.4]{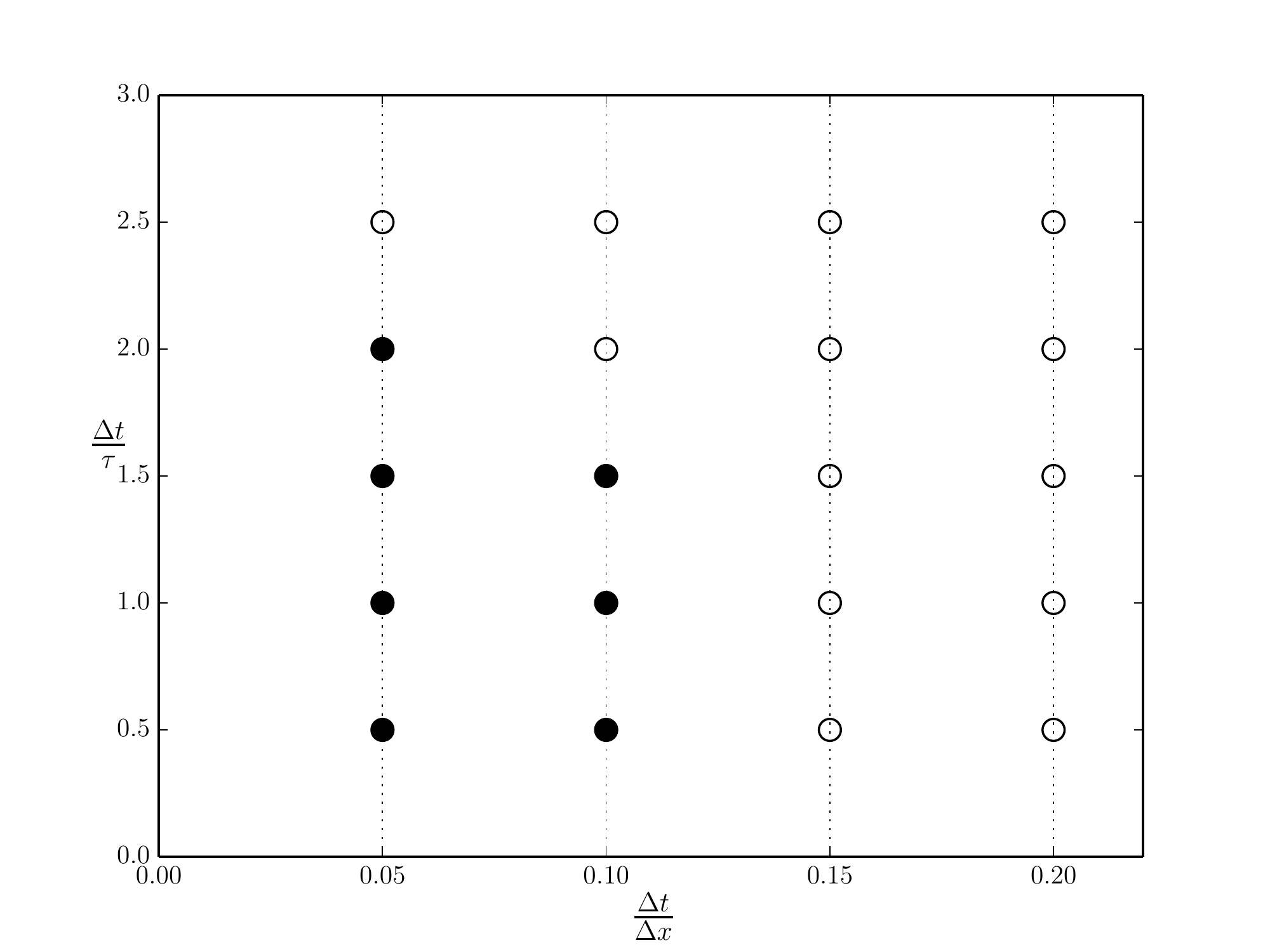}
\par\end{center}

\caption{Stability region for the RK2 scheme at $Re=100$.\label{fig:Stab_RK2}}
\end{minipage}\hfill{}%
\begin{minipage}[c][1\totalheight][t]{0.45\textwidth}%
\begin{center}
\includegraphics[scale=0.4]{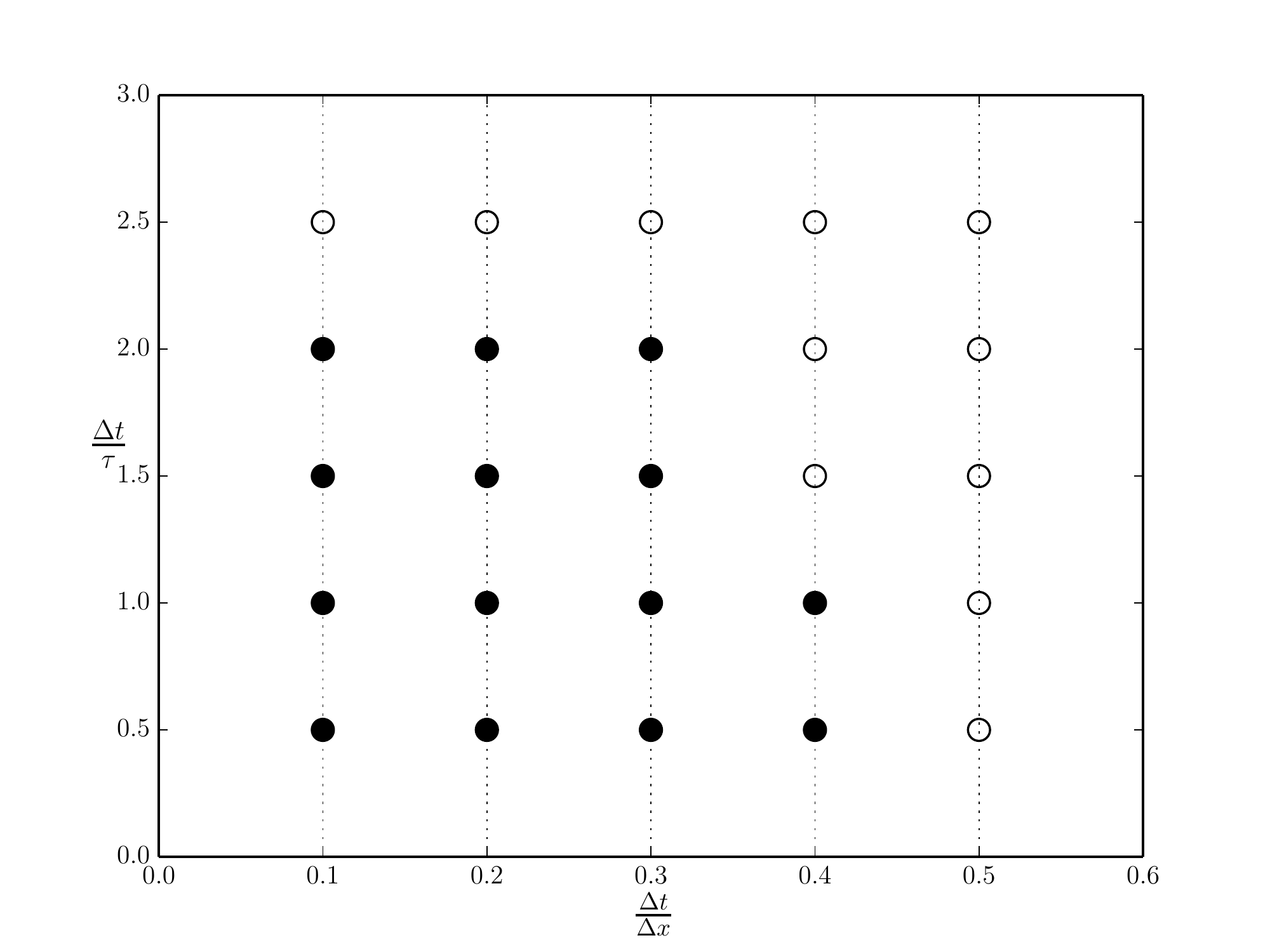}
\par\end{center}

\caption{Stability region for the GZ scheme at $Re=100$.\label{fig:Stab_Guo}}
\end{minipage}
\end{figure}
 
\par\end{center}

From Figure \ref{fig:Stab_RK2}, we can observe that the RK2-based
scheme is unstable beyond $\Delta t/\tau>2$, irrespective of the
$CFL$ number. Moreover, the stable solutions which are obtained at
$\Delta t/\tau<2$ are done so at very low $CFL$ values. As described
before, the small $\Delta t$ requirement stems from the fact that
as $Re$ increases, the collision term becomes more non-linear and
stiff, which requires an implicit approximation for numerical stability.
Stability of RK2-based schemes can be increased slightly by adopting
second-order upwind schemes; however, the marginal stability is added
at the cost of increasing numerical diffusion. Similarly, multi-stage
Runge-Kutta schemes such as RK4 can also increase stability. However,
all RK-based OLB schemes involve multiple evaluation of the $f_{i}^{eq}$
term for each advancement in $\Delta t$, the number of evaluations
depending upon the stages in the scheme. Since evaluation of $f^{eq}$
is a computationally intensive step, RK-based schemes are also computationally
inefficient. Therefore, we can conclude that RK2 based OLB schemes
are not particularly suitable for flows with large non-linearities. 

Figure \ref{fig:Stab_Guo} shows the stability region for the GZ scheme
for $Re=100$. From the stability-region plot, we can again see that
the scheme is unstable for all $\Delta t/\tau>2$, irrespective of
the $CFL$ number. Thus, in spite of the implicit collision approximation,
large $\Delta t$ could not be used in the scheme. This observation
is consistent with the stability analysis presented by the authors
of the scheme.

The stability regions of the AE/CE scheme are shown in Figure \ref{fig:Stab_AECE_400}
and \ref{fig:Stab_AECE_1000}. From the figures, it is evident that
the AE/CE scheme does allow $\Delta t/\tau>2$, but does so only at
lower $CFL$ number $(CFL<0.2)$. As $Re$ increases, however, the
explicit collision approximation becomes inadequate, and hence the
$\Delta t$ has to be smaller. 

\begin{center}
\begin{figure}[H]
\begin{minipage}[c][1\totalheight][t]{0.45\textwidth}%
\begin{center}
\includegraphics[scale=0.4]{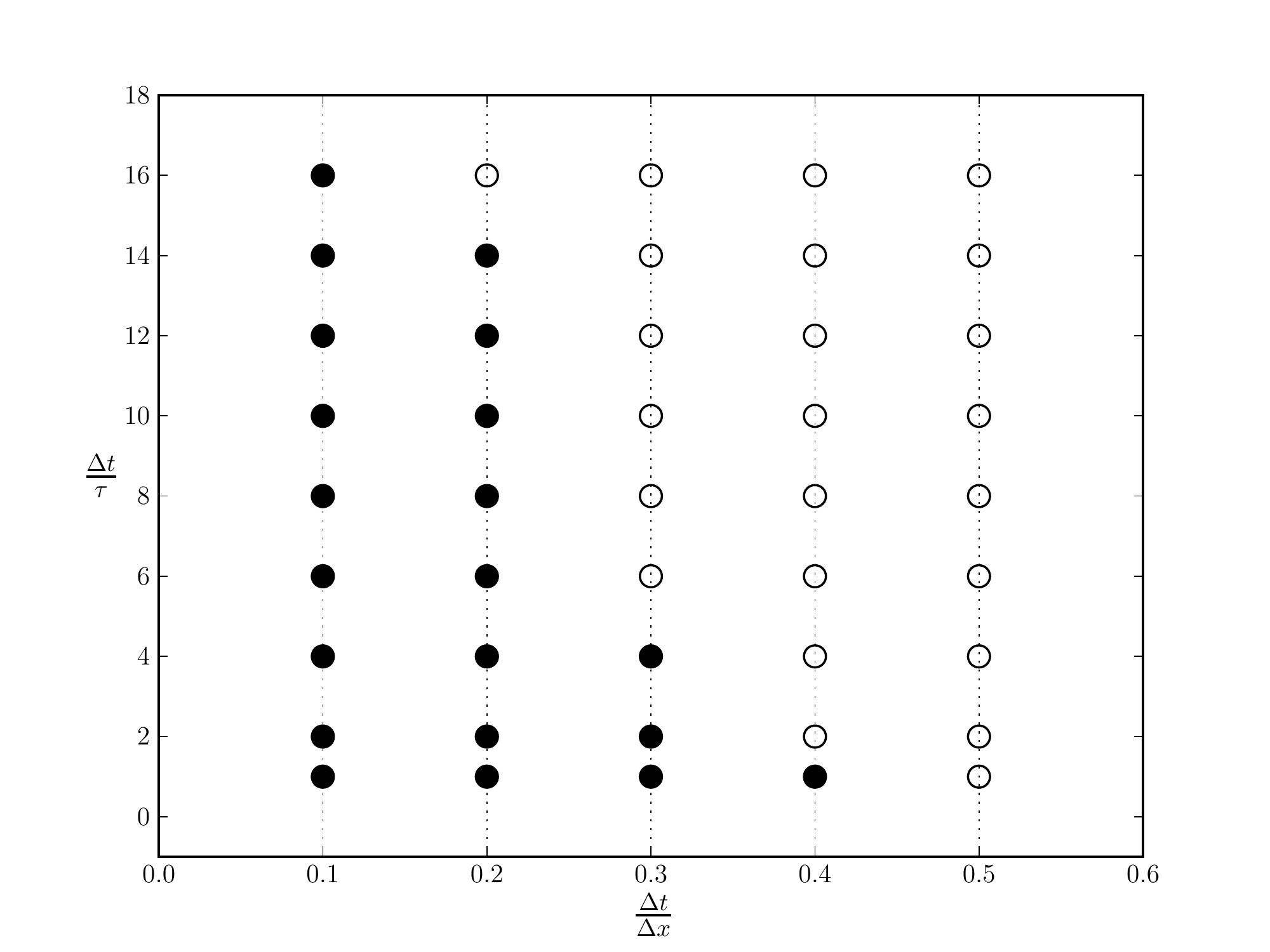}
\par\end{center}

\caption{Stability region for the AE/CE scheme at $Re=400$.\label{fig:Stab_AECE_400}}
\end{minipage}\hfill{}%
\begin{minipage}[c][1\totalheight][t]{0.45\textwidth}%
\begin{center}
\includegraphics[scale=0.4]{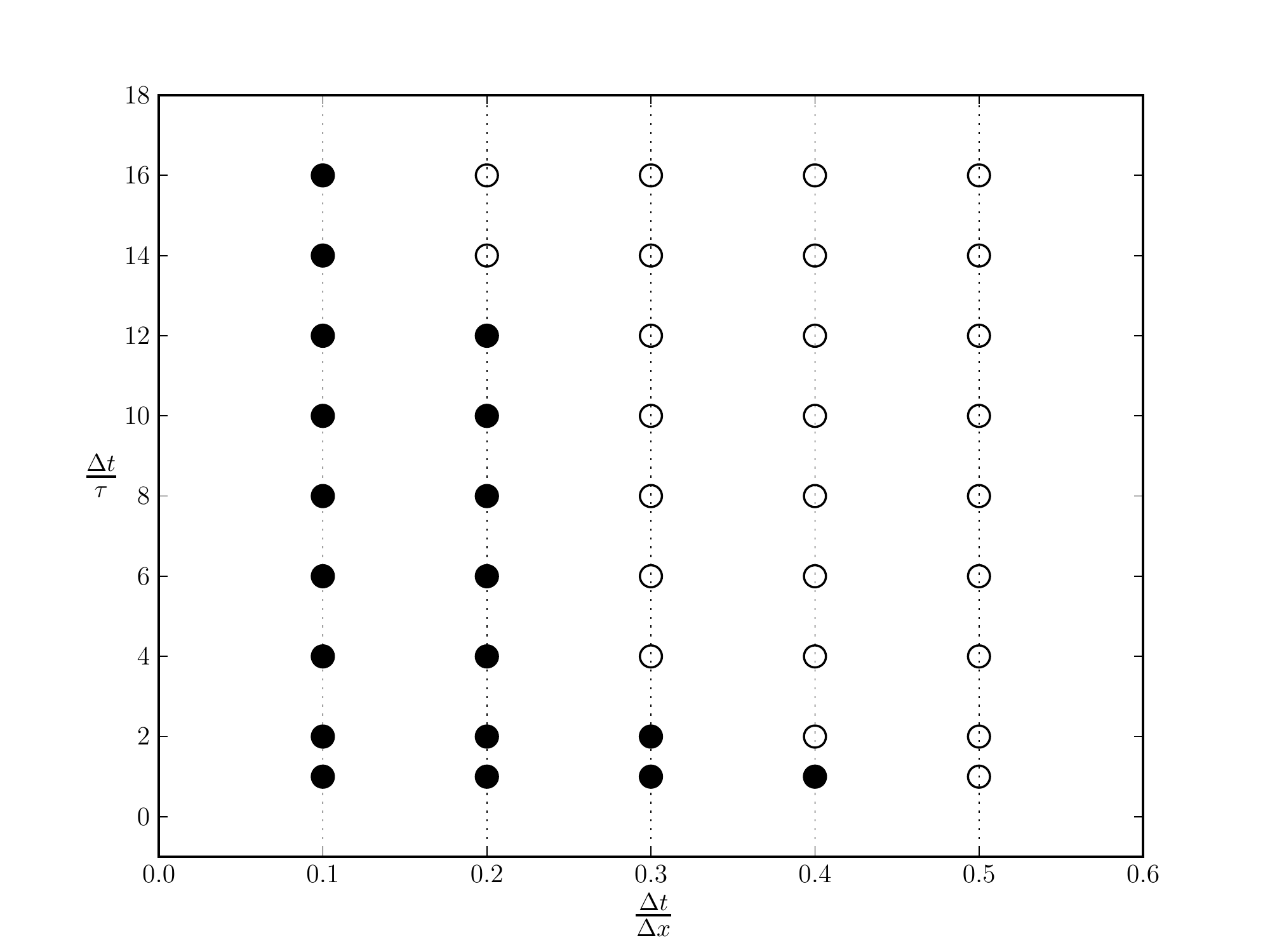}
\par\end{center}

\caption{Stability region for the AE/CE scheme at $Re=1000$.\label{fig:Stab_AECE_1000}}
\end{minipage}
\end{figure}

\par\end{center}

Finally, Figures \ref{fig:Stab_Bardow_400} and \ref{fig:Stab_Bard_1000}
show the stability regions for the BKG scheme at $Re=400$ and $1000$.
We can observe that the simulations are stable at $\Delta t/\tau$
as high as 30, and that at high $CFL$ numbers. The trend also does
not deteriorate with increasing $Re$ in the range that we have tested.
This demonstrates the unconditional collision stability of the BKG
scheme, with $\Delta t$ restricted only by the local $CFL$ number
due to explicit advection. 

The BKG scheme is also computationally more \emph{efficient} than
the comparable AE/CI scheme. This is because whereas a predictor-corrector
type of scheme is needed for the implicit collision approximation
in the AE/CI scheme, a simple variable transformation given by Equation
\ref{eq:variable transf.} is required in the BKG scheme. Additionally,
the BKG scheme, as with all advection-explicit OLB schemes, retains
the data-locality feature of the LB method, and hence can be easily
parallelized. 

\textcolor{blue}{Finally, it is important to note that we have discussed
only the numerical stability of different OLB schemes in terms of
maximum allowable $\Delta t.$ However, as for all explicit advection
schemes, including the BKG scheme, much smaller values of $\Delta\boldsymbol{x}$
and $\Delta t$ are required to get accurate solutions. }

\begin{center}
\begin{figure}[H]
\begin{minipage}[c][1\totalheight][t]{0.45\textwidth}%
\begin{center}
\includegraphics[scale=0.4]{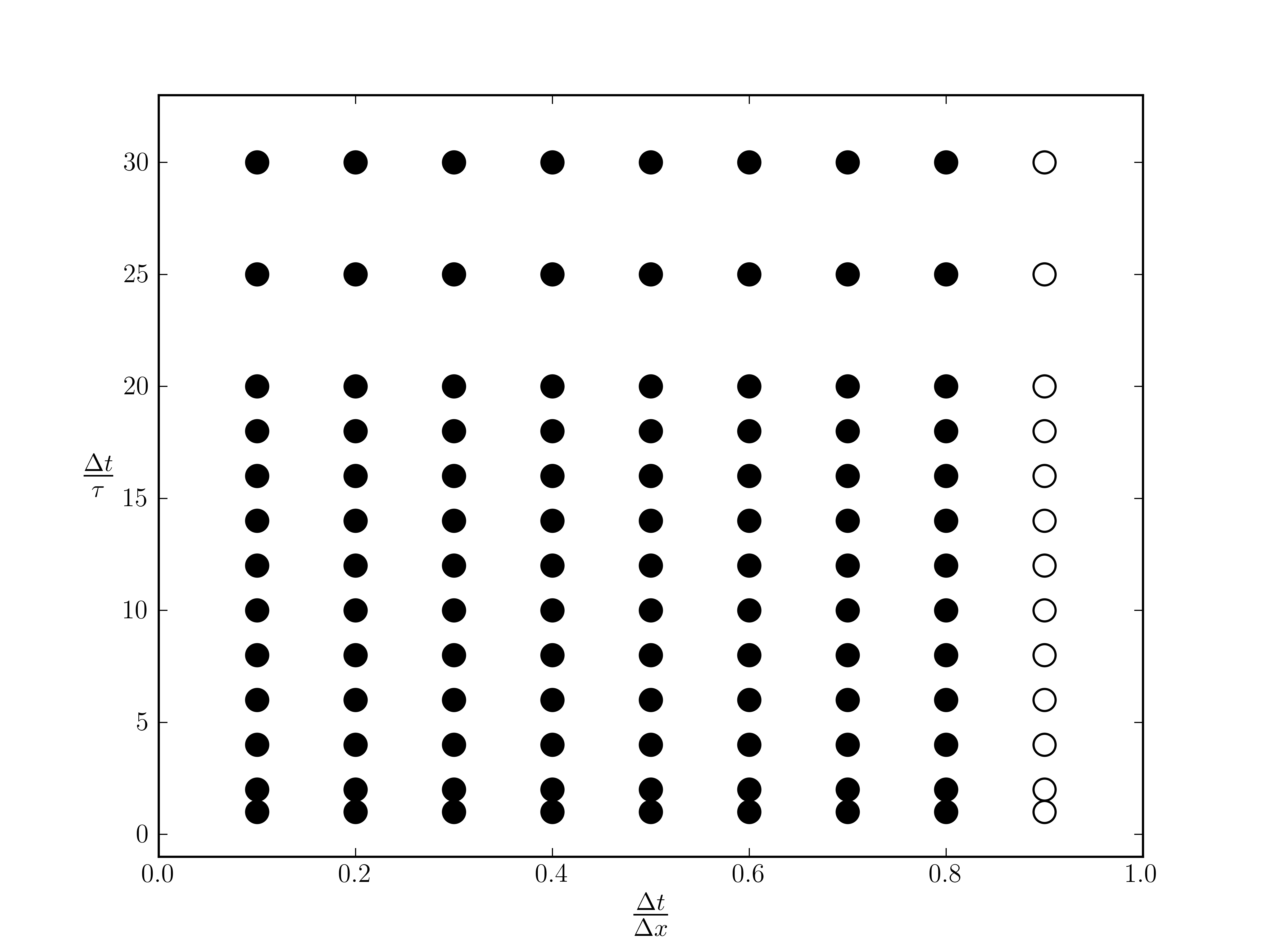}
\par\end{center}

\caption{Stability region for the BKG scheme at $Re=400$.\label{fig:Stab_Bardow_400}}
\end{minipage}\hfill{}%
\begin{minipage}[c][1\totalheight][t]{0.45\textwidth}%
\begin{center}
\includegraphics[scale=0.4]{Figures/Bardow_stabilityRe100_1000}
\par\end{center}

\caption{Stability region for the BKG scheme at $Re=1000$.\label{fig:Stab_Bard_1000}}
\end{minipage}
\end{figure}

\par\end{center}

\section{Conclusions}

In this work, several explicit OLB schemes have been compared by implementing
them for benchmark flow problems. The following conclusions can be
drawn from the observations:
\begin{enumerate}
\item The characteristics-based OLB schemes provide higher numerical stability
compared to RK-based schemes.
\item In characteristics based-schemes, the $\Delta t<\tau$ constraint
no longer applies, even at high $Re$. 
\item The scheme proposed by Bardow \emph{et al.} is stable over a much
wider range of simulation parameters $(\Delta t/\tau,\,\Delta t/\Delta x)$
compared to other similar characteristics-based OLB schemes for the
problems tested in this work. The BKG scheme also retains the simple
explicit form of the LB method, while providing unconditional collision-stability. 
\end{enumerate}
These conclusions indicate that the BKG scheme provides the most stable
and efficient explicit time-marching scheme, which can be extended
to flow problems with FV or FD discretization. This scheme, in theory,
can also be adopted for thermal problems with both off- and on-lattice
discrete-velocity sets \citep{Bardow2008}.

\subsection*{Acknowledgment}

This material is based upon work supported by the National Science
Foundation under grant no. CBET-1233106.

\section*{References}

\bibliographystyle{elsarticle-num}
\bibliography{\string"LBM_reference _database\string"}

\end{document}